\documentclass[notitlepage,superscriptaddress,nofootinbib,tightenlines,preprintnumbers,twocolumn,longbibliography]{revtex4-2}
\usepackage{amsmath,verbatim,latexsym,amssymb,indentfirst,mathrsfs,mathtools,amsthm,bbm,bm,hyperref,url,cancel,subcaption,enumitem}
\usepackage[font=small,labelfont=bf,format=plain,justification=raggedright,singlelinecheck=false]{caption}
\usepackage{verbatim,indentfirst}
\usepackage[title,titletoc]{appendix}

\usepackage{graphicx}
\usepackage{epstopdf}
\newcommand{\caphead}[1]{{\bf #1}}

\renewcommand{\thesection}{\Roman{section}}
\renewcommand{\thesubsection}{\Alph{subsection}}
\renewcommand{\thesubsubsection}{\Alph{subsection} \arabic{subsubsection}}
\makeatletter
\def\p@subsection{}
\makeatother
\makeatletter
\def\p@subsubsection{}
\makeatother

\makeatletter
\newcommand\footnoteref[1]{\protected@xdef\@thefnmark{\ref{#1}}\@footnotemark}
\makeatother

\usepackage{soul}

\usepackage{color}
\usepackage[dvipsnames]{xcolor}
\usepackage[normalem]{ulem}

\newcommand{\hot}{\mathcal{H}}
\newcommand{\cold}{\mathcal{C}}
\newcommand{\tar}{\mathcal{T}}
\newcommand{\switch}{\mathcal{S}}
\newcommand{\rest}{\mathcal{R}}
\newcommand{\ladder}{\mathcal{L}}
\newcommand{\pure}{\mathcal{P}}
\newcommand{\dead}{{\rm D}}

\newcommand{\A}{\mathcal{A}}
\newcommand{\B}{\mathcal{B}}

\newcommand{\inter}{ {\rm int} }   

\newcommand{\tot}{ {\rm tot} }
\newcommand{\Tr}{{\rm Tr}}   
\def\id{\mathbbm{1}}   
\newcommand{\kB}{k_\mathrm{B}}  

\newcommand{\Hil}{\mathscr{H}}  
\newcommand{\Sys}{\mathcal{S}}  
\newcommand{\Env}{\mathcal{E}}  
\newcommand{\Dim}{d}   

\newcommand{\LParen}{ \bm{(} }
\newcommand{\RParen}{ \bm{)} }

\renewcommand\th{ {\rm th} }

\newcommand*{\bra}[1]{\langle #1\rvert}
\newcommand*{\ket}[1]{\lvert #1 \rangle}

\newcommand*{\ketbra}[2]{\lvert #1 \rangle\!\langle #2 \rvert}

\begin{document}
 
\title{Key Issues Review: Useful autonomous quantum machines}
\author{Jos\'{e} Antonio Mar\'{i}n Guzm\'{a}n}
\affiliation{Joint Center for Quantum Information and Computer Science, NIST and University of Maryland, College Park, MD 20742, USA}
\author{Paul Erker}
\affiliation{Atominstitut, Technische Universit\"{a}t Wien, 1020 Vienna, Austria}
\affiliation{Institute for Quantum Optics and Quantum Information (IQOQI),
Austrian Academy of Sciences, Boltzmanngasse 3, 1090 Vienna, Austria}
\author{Simone Gasparinetti}
\affiliation{Department of Microtechnology and Nanoscience, Chalmers University of Technology, 412 96 Gothenburg, Sweden
}
\author{Marcus Huber}
\affiliation{Atominstitut, Technische Universit\"{a}t Wien, 1020 Vienna, Austria}
\affiliation{Institute for Quantum Optics and Quantum Information (IQOQI),
Austrian Academy of Sciences, Boltzmanngasse 3, 1090 Vienna, Austria}
\author{Nicole~Yunger~Halpern}
\email{nicoleyh@umd.edu}
\affiliation{Joint Center for Quantum Information and Computer Science, NIST and University of Maryland, College Park, MD 20742, USA}
\affiliation{Institute for Physical Science and Technology, University of Maryland, College Park, MD 20742, USA}
\date{\today}

%
%
\begin{abstract}
Controlled quantum machines have matured significantly. A natural next step is to increasingly grant them autonomy, freeing them from time-dependent external control. For example, autonomy could pare down the classical control wires that heat and decohere quantum circuits; and an autonomous quantum refrigerator recently reset a superconducting qubit to near its ground state, as is necessary before a computation. Which fundamental conditions are necessary for realizing useful autonomous quantum machines? Inspired by recent quantum thermodynamics and chemistry, we posit conditions analogous to DiVincenzo's criteria for quantum computing. Furthermore, we illustrate the criteria with multiple autonomous quantum machines (refrigerators, circuits, clocks, etc.) and multiple candidate platforms (neutral atoms, molecules, superconducting qubits, etc.). Our criteria are intended to foment and guide the development of useful autonomous quantum machines.
\end{abstract}

{\let\newpage\relax\maketitle}

\begin{figure}[hbt]
\centering
\includegraphics[width=.25\textwidth, clip=true]{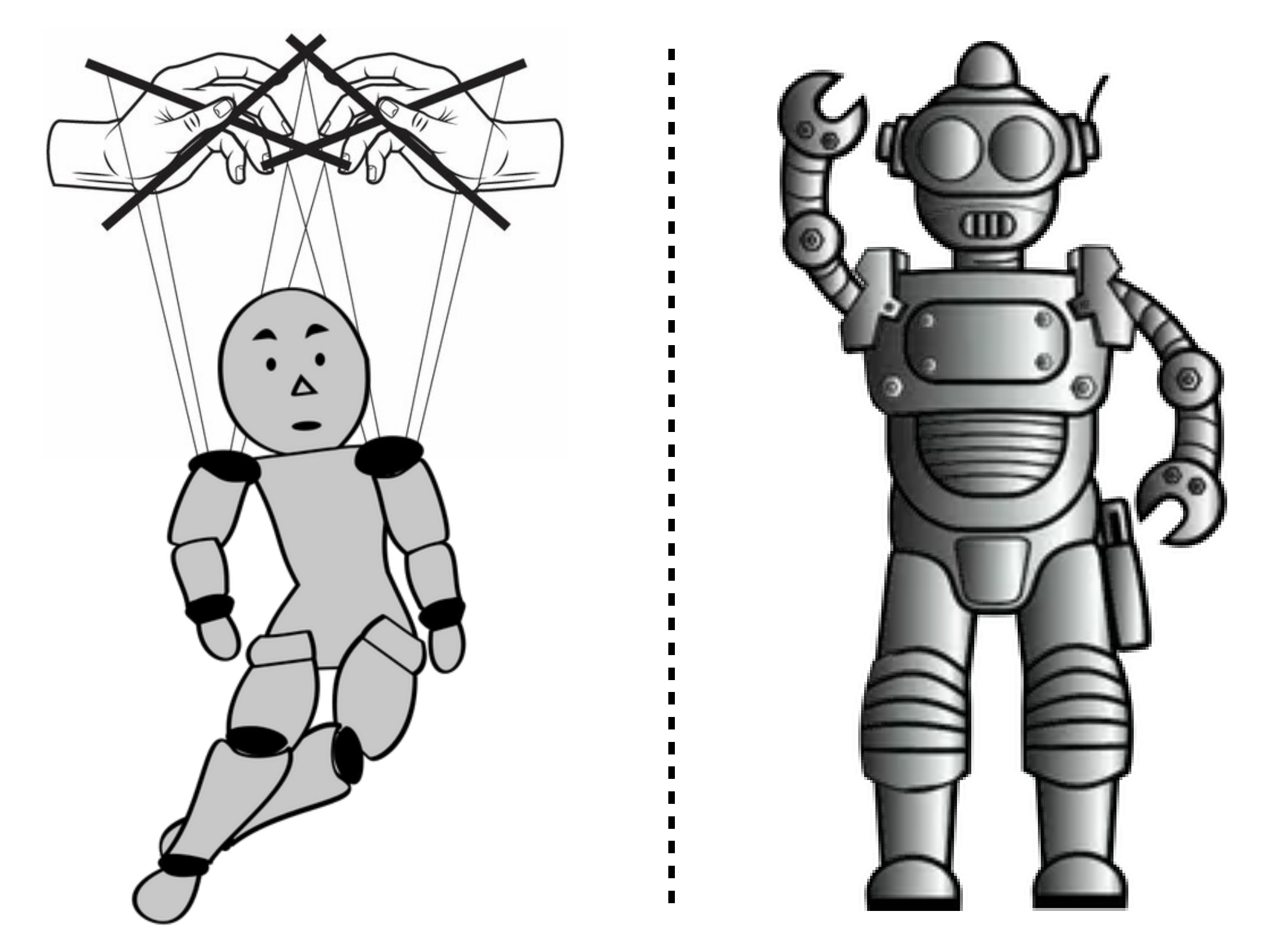}
\caption{\caphead{Nonautonomous vs. autonomous machines:} The hands apply time-dependent external control to the nonautonomous machine (left). The autonomous machine (right) evolves under a constant Hamiltonian. This machine may carry a power source or draw its own free energy (or a nonequilibrium generalization thereof) from its environment.}
\label{fig_Illustrate_AM}
\end{figure}

Automata are machines that operate independently of external control (Fig.~\ref{fig_Illustrate_AM}). In an early example, the ancient Greek mathematician Archytas of Tarentum supposedly built a wooden pigeon powered by steam~\cite{Archytas_17}.
For centuries, automata served as curiosities for entertainment and for impressing visitors. Not until the 18th-century Industrial Revolution did engineers harness automata for practical purposes on large scales. (We use an expansive definition of \emph{automaton}, not restricting the term to machines that resemble humans or animals.) Today, automata speed up manufacturing, deliver packages, drive car passengers, and even clean kitchen floors. In summary, classical automata have progressed from curios to useful tools.

Autonomous quantum machines have embarked upon a progression that we hope will end analogously. 
Quantum thermodynamicists have designed theoretical autonomous engines~\cite{Scovil_59_Three,Geusic_67_Quantum,Youssef_10_Quantum,Sanchez_11_Optimal,Gilz_13_Generalized,GelbwaserKlimovsky_13_Work,Mari_15_Quantum,Gelbwaser_15_Work,Roche_15_Harvesting,Alicki_17_Thermodynamic,Roulet_17_Autonomous,Seah_18_Work,Roulet_18_Autonomous,Roulet_18_Autonomous,Fogedby_18_Autonomous,Hammam_21_Optimizing,Mayrhofer_21_Stochastic,Niedenzu_19_Concepts,Wachtler_19_Proposal,Drewsen_19_Quantum,Verteletsky_20_Revealing,Strasberg_21_Autonomous,RignonBret_21_Thermodynamics,Opatrny_23_Nonlinear}, refrigerators~\cite{linden2010, levy2012,  chen2012g, venturelli2013, correa2014, silva2015, Mitchison_15_Coherence, hofer2016a, Silva_16_Performance, Mitchison_16_Realising, mu2017, nimmrichter2017, du2018, mitchison2018, Mukhopadhyay_18_Quantum, Erdman_18_Absorption,holubec2019, Manzano_19_Boosting, Das_19_Necessarily, Maslennikov_19_Quantum, naseem2020, Hewgill_20_Three, manikandan2020, arrangoiz_arriola2018, bhandari2021, kloc2021,Mayrhofer_21_Stochastic, almasri2022, Okane_22_Quantum, BohrBrask_22_Small, AbdouChakour_22_Coupling, Mohanta_22_Universal}, clocks~\cite{Erker_17_Autonomous,Woods_19_Autonomous,Schwarzhans_21_Autonomous,Woods_21_Autonomous,Woods_22_Quantum,Woods_23_Autonomous,Manikandan_23_Autonomous,Manikandan_23_Autonomous}, 
Maxwell demons~\cite{Strasberg_13_Thermodynamics,Strasberg_18_Fermionic,Koski_15_On,Sanchez_19_Autonomous}, and more~\cite{BohrBrask_15_Autonomous,Manzano_15_Autonomous,Chiranjib_18_Generating,Monsel_18_Arrow,Latune_19_Quantum,BohrBrask_22_Operational} without external drives~\cite{Tonner_05_Autonomous,mitchison2019}. Some of these machines do not even require thermodynamic-work inputs. Instead, the refrigerators and clocks siphon off heat flowing between different-temperature baths nearby. 
Most of these studies explore foundational matters: whether one can build autonomous quantum machines in principle, fundamental limits on such a machine's performance, etc. This work parallels Archytas and his successors as they sketched designs, toyed with steam propulsion, and so on.

%

In a next step along the hoped-for progression, experimentalists have just begun building autonomous quantum machines. Platforms used include trapped ions~\cite{Maslennikov_19_Quantum}, superconducting qubits~\cite{Ali_23_Thermally}, molecules~\cite{Kucharski_14_Templated}, single-electron boxes~\cite{Koski_15_On}, and quantum dots~\cite{Hartmann_15_Voltage,Josefsson_18_Quantum}.
For instance, trapped ions and superconducting qubits have realized autonomous quantum refrigerators~\cite{Maslennikov_19_Quantum,Ali_23_Thermally}. Such experiments have initiated the bridge from theory to reality. Yet they largely resemble Archytas's pigeon: They are impressive curiosities, not practical tools. 

A quantum engine offers an example. Conventional thermodynamics spotlights engines formed from classical gases. An early question in quantum thermodynamics was~\cite{Kosloff_14_Quantum,Vinjanampathy_16_Quantum,mitchison2019} \emph{can quantum systems similarly perform work as engines?} The answer is yes: quantum Carnot, Otto, and Stirling engine cycles have been defined. So have continuous quantum engines, which need not be cycled through finite-time strokes~\cite{Kosloff_14_Quantum}.\footnote{
Quantum engines differ from many biological engines, or molecular motors~\cite{Iino_20_Introduction}. Molecular motors tend to behave classically. Examples include the motor that moves a bacterium's flagellum.}
Some such quantum engines have been realized experimentally. Examples include an engine formed from one natural atom, as in~\cite{VanHorne_20_Single}. This experiment is laudable for demonstrating quantum control and for extending thermodynamic principles to the quantum regime. Nevertheless, the engine cannot earn its keep. The engine's ``working fluid'' consists of two energy levels separated by an optical transition. Therefore, one may expect to extract $\approx 1$ eV of work per cycle.
Yet preparing the atom required Doppler and sideband cooling---far more work.\footnote{
The engine was not autonomous, so operating the engine required more work inputs. However, we focus on the work inputs that the engine would have required if autonomous.}
Hence the engine is impractical.

One exception was initiated recently: the autonomous quantum refrigerator formed from superconducting qubits~\cite{Ali_23_Thermally}. The quantum refrigerator cooled a target qubit to below the temperatures realizable via passive thermalization with the dilution-refrigerator environment. This target qubit, reset to near its ground state, could potentially serve in a later quantum computation. In such an application, the quantum refrigerator could draw energy from the temperature gradient between the dilution refrigerator's inner and outer plates. Cooling this autonomous refrigerator to the quantum regime would cost negligible extra work, the dilution refrigerator already being cold to support the upcoming computation. Work is required to extend the proof-of-principle experiment~\cite{Ali_23_Thermally} to applications. Nevertheless, the experiment demonstrates the autonomous quantum refrigerator's potential for usefulness. Experimentalists have cooled superconducting qubits alternatively via active reset~\cite{Riste_12_Initialization,Tholen_22_Measurement,Salathe_18_Low,Moreira_23_Realization} and via other unconditional-reset protocols~\cite{Magnard_18_Fast,Zhou_21_Rapid}. One can attempt to compare all the strategies through their reset times and the targets' late-time excited-state populations. The results are mixed: the quantum refrigerator and competitors outperform each other in different ways. Still, in some situations, the autonomous quantum refrigerator's lesser need for control may outweigh any benefits enjoyed by competitors.

One can envision other useful autonomous quantum machines, five types of which we sketch here in motivating this Perspective.\footnote{
Practicality has motivated the design of nonautonomous quantum thermal machines~\cite{manikandan2023a, Karmakar_22_Cyclic,baugh2005, solfanelli2022, buffoni2023}, too.}
These machines, with the refrigerator mentioned above, serve as illustrative examples. Criteria are general, abstract concepts and so require examples to provide grounding and concreteness. This list, although extensive, is intended to provide a brief survey, as suits a review's introduction, rather than a thorough tour of all conceivable autonomous quantum machines. To illustrate the range of possible machines, we progress from experimentally realized ones to natural ones, to as-yet imaginary machines to which the community can aspire:
\begin{enumerate}

   \item Autonomous quantum circuits would apply their own gates, offering several possible benefits.\footnote{
\emph{Autonomous quantum computing} is not synonymous with \emph{quantum machine learning}. During the latter, an input state undergoes quantum gates that may be implemented via time-dependent external control. For instance, some platforms undergo single-qubit gates when an external field is switched on and off. Such time-dependent external control is incompatible with autonomous operation. However, quantum machine learning might be rendered autonomous, in our sense of the term, similarly to ordinary quantum computing.}
Autonomy could pare down the control wires in superconducting-qubit architectures. Control wires limit the number of superconducting qubits that fit on a chip~\cite[App. C.4.I]{Grumbling_19_Quantum}. Also, control wires are dissipative macroscopic, classical objects that heat qubits up~\cite[Sec.~2.1]{Krinner_19_Engineering}.
Eliminating control wires could therefore benefit superconducting-qubit circuits' scalability and coherence times. Similarly, quantum dots are controlled by gate voltages---classical control whose parameter space has grown unwieldy~\cite{Zwolak_23_Advances}. Hence autonomy could improve also semiconductor quantum circuits' scalability.

   \item To apply gates at the proper times, autonomous quantum circuits would need autonomous quantum machines of a second type: clocks. Theorists have designed autonomous quantum clocks that tick by emitting photons~\cite{Erker_17_Autonomous}. Such clocks differ from the quantum clocks used today, formed from ultracold atoms interacting with lasers~\cite{Campbell_11_Ultracold}: Today's quantum clocks receive external feedback when the atoms are measured, the lasers are tuned, etc.

   \item Autonomous quantum machines of a third type 
could detect, transduce, and transmit energy.
Nature has produced machines of this type.
\emph{Photoisomers} are molecules operative across nature and technologies~\cite{Waldeck_91_Photoisomerization,Bandara_12_Photoisomerization,Hahn_00_Quantum,Browne_06_Making}. Examples include retinal, found in the protein rhodopsin in our eyes~\cite{Strauss_05_Retinal}. A photoisomer has one shape in thermal equilibrium at biological temperatures. Absorbing a photon---say, from the sun---enables the molecule to switch configurations. The switching can galvanize chemical reactions leading to the experience of sight. Photoisomers exhibit quantum phenomena including coherence relative to the energy eigenbasis, as well as nonadiabatic evolution.
Beyond photoisomers, chemistry features other autonomous quantum machines: 
Photosynthetic complexes transform light into chemical energy~\cite{Li_23_Single}, enzymes interconvert molecules with help from tunneling~\cite{Klinman_13_Hydrogen}, etc.

   \item One can imagine granting quantum sensors autonomy~\cite{NLG_24_Do}. 
Quantum sensors detect small magnetic fields and temperature gradients~\cite{Degen_17_Quantum}. 
Realizations include nitrogen-vacancy centers in diamond, neutral atoms, trapped ions, superconducting qubits, optomechanical oscillators, and more~\cite[Sec.~III]{Degen_17_Quantum}.
External control includes microwave pulses and placement near the to-be-detected source.
For instance, experimentalists have injected nanodiamond sensors into embryo tissues whose temperatures need measuring~\cite{Choi_20_Probing,Masazumi_20_Real}. 
One can envision quantum sensors that reach and report about sources autonomously.
For example, a photoisomer---a natural autonomous quantum photodetector---might follow a potential gradient in its environment until reaching a light source. Alternatively, functional groups can be attached to larger quantum objects, such as molecules, and have been proposed in sensing applications~\cite{Zhu_22_Functionalizing}. The molecule could serve as a vehicle for transporting the functional-group sensor.

   \item In a fifth illustration of possible autonomous quantum machines, we let our imaginations loose. One can envisage autonomous quantum machines assembling bespoke molecules, servicing quantum computers, delivering atoms as drones, or building other quantum devices. Such machines may seem like castles in the air. But so did controlled quantum computers, decades ago; and quantum computers have been built and are being scaled up. DiVincenzo's criteria have guided quantum computers from castle-in-the-air status to reality~\cite{DiVincenzo_00_Physical}. He posited five criteria necessary for building a quantum computer, plus two optional criteria necessary for information transmission. 
\end{enumerate}

Analogously, we posit eight criteria necessary for building a useful autonomous quantum machine. Two optional criteria concern transportation and information transmission. 
We devised these criteria by thinking fundamentally about what autonomous quantum machines need and do, as well as by abstracting general principles from example machines in the literature. Our criteria therefore concern a general autonomous quantum machine, regardless of its task (work extraction, cooling, timekeeping, etc.). However, we illustrate our criteria with example machines that undertake specific tasks, including the example machines mentioned above (engines, refrigerators, clocks, etc.). Also, we illustrate with various possible platforms (superconducting qubits, neutral atoms, etc.). This approach parallels DiVincenzo's: he could not detail his criteria's realizations in all possible quantum-computing platforms. Therefore, he posited general principles and illustrated them.
We hope that our criteria, analogously to DiVincenzo's, guide experimentalists in realizing useful autonomous machines. Before presenting our criteria, we stipulate what we mean by \emph{machine}, \emph{autonomous}, and \emph{quantum}. The criteria appear in~\ref{sec_Criteria}. Section~\ref{sec_Outlook} details practical challenges and possible solutions to them.

\section{Definitions}
\label{sec_Def}

Before presenting our criteria, we specify meanings for \emph{machine}, \emph{autonomous}, and \emph{quantum}. 
Alternative definitions may exist, as the terms are broad. Still, we believe our definitions to be reasonable. Furthermore, specifying them will clarify our criteria:
\begin{enumerate}[label=(\alph*)]

   \item \emph{\textbf{Machine:} A physical device, potentially formed from components working together, that harnesses energy to accomplish a task.}
   
   \item \emph{\textbf{Autonomous:} A machine is autonomous if its total microscopic Hamiltonian is not changed by any agent using the machine.
   } 
   An autonomous machine's user performs no thermodynamic work on the machine during the machine's operation. Work tends to be defined through changes in the system-of-interest Hamiltonian, in quantum thermodynamics~\cite{Vinjanampathy_16_Quantum}.\footnote{
   For example, an adenosine-triphosphate molecule provides chemical work via hydrolyzation: one of its phosphates splits off from the rest of the molecule, which becomes adenosine diphosphate. The cleaving of the bond releases energy and changes the molecule's Hamiltonian.}
   
   Microscopic Hamiltonians contrast with effective Hamiltonians. An effective Hamiltonian can result from shifting a microscopic Hamiltonian into a rotating reference frame, then dropping small terms. A microscopic Hamiltonian remains constant in the absence of time-dependent external drives. 
   
   \item \emph{\textbf{Quantum:} A system is quantum if the axioms of quantum theory describe the system usefully.} Quantum systems can, but need not, exhibit quantum phenomena such as entanglement, coherence relative to relevant bases, discretized spectra, measurement disturbance, contextuality~\cite{Bell_66_On,Kochen_67_Problem,Spekkens_05_Contextuality}, and the quantum-computational resource called magic~\cite{Howard_14_Contextuality}. Classical systems can approximate discretized spectra, and classical waves exhibit coherence. Quantum theory's axioms describe classical systems, but not usefully; classical theories offer greater calculational efficiency and physical insight. For example, quantum theory ultimately models your toothbrush. However, most physicists would recommend Newtonian mechanics for describing the trajectory followed by a toothbrush as it falls. We do not mean \emph{nonclassical} by \emph{quantum}. (One might call a phenomenon \emph{nonclassical} if, for example, no noncontextual ontological model~\cite{Bell_66_On,Kochen_67_Problem,Spekkens_05_Contextuality}, often regarded as a classical theory, reproduces the phenomenon.)
   
   Relatedly, this Perspective does not posit criteria under which autonomous quantum machines generally outperform all classical counterparts. Rather, the Perspective provides criteria under which autonomous quantum machines are useful. We believe this goal to be worthwhile, ambitious, and reasonable for the near future. Farther in the future, another Perspective may posit criteria under which autonomous quantum machines generally beat all classical competitors.

\end{enumerate}

\section{Criteria}
\label{sec_Criteria}

This section contains our DiVincenzo-like criteria for realizing useful autonomous quantum machines. Table~\ref{tab_Criteria} summarizes the criteria. We denote by $\sigma_a$ the Pauli-$a$ operator, for $a = x, y, z$; by $\ket{1}$, the eigenvalue-1 eigenstate of $\sigma_z$; and by $\ket{0}$, the eigenvalue-$(-1)$ eigenstate.

%
%
\begin{table*}[t]
\begin{center}
\begin{tabular}{|c|c|}
   \hline  Subsection  &  Criterion  \\  \hline  \hline
   A  &  Access to useful energy   \\    \hline
   B  &  Processing unit or target  \\  \hline
   C  &  Interactions among the machine's components \\ \hline
   D  &  Timekeeping mechanism  \\  \hline
   E  &  Structural integrity  \\  \hline
   F  &  Sufficient purity  \\  \hline
   G  &  Output worth the input  \\  \hline
   H  &  Ability to switch off after completing assignment  \\  \hline
   I  &  Mobility (optional)  \\  \hline
   J  &  Interoperability (optional)
   \\    \hline
\end{tabular}
\caption{\caphead{Summary: DiVincenzo-like criteria for autonomous quantum machines.}}
\label{tab_Criteria}
\end{center}
\end{table*}
\subsection{Access to useful energy}
\label{crit_Energy}

Useful energy enables one to perform thermodynamic work. Work empowers a machine to direct its motion---to overcome its momentum and random buffets from its environment. Free energy, or a nonequilibrium generalization thereof, offers the capacity to perform work. A machine can access this energy directly or indirectly, as we now discuss. 

We label as a \emph{battery} any system that reliably stores (the capacity to perform) work and from which work can reliably be retrieved. Small-scale batteries include adenosine triphosphate (ATP), a molecule that powers chemical reactions in cells. Autonomous classical nanowalkers leverage ATP, as discussed under criterion~\ref{crit_Mobility}. Quantum thermodynamics features multiple battery models (e.g.,~\cite{Horodecki_13_Fundamental,Brandao_15_Second,Skrzypczyk_13_Extracting,NYH_16_Beyond,NYH_18_Beyond,Lyu_23_Efficient,Binder_15_Quantacell}), many idealized. Examples include a \emph{work bit}, a two-level system governed by a Hamiltonian $\Delta \sigma_z$, for $\Delta > 0$~\cite{Horodecki_13_Fundamental,Brandao_15_Second}. 
A work bit charges during $\ket{0}  \mapsto  \ket{1}$
and provides work during
$\ket{1}  \mapsto  \ket{0}$. 

Other energy sources 
provide work indirectly. They are nonequilibrium systems that contain free energy (or a generalization thereof), which a machine can harvest to produce work.
Often, such systems have gradients of temperature, chemical potential, or other generalized thermodynamic forces. The commonest example consists of two heat baths. One, at a temperature $T_\cold$, is out of equilibrium with a heat bath at a temperature $T_\hot > T_\cold$. Heat flows between the baths. 
This heat is not work, so the temperature gradient does not provide work directly. However,
a machine can siphon off a fraction of the current 
and transform it partially (in accordance with Carnot's theorem) into work.
Such machines are called \emph{absorption machines}~\cite{mitchison2019}. Heat baths of two types have fueled superconducting-qubit machines: resonators~\cite{Ronzani_18_Tunable} and thermal fields propagating through microwave waveguides~\cite{Ali_23_Thermally}.
Despite providing energy, baths can threaten the purity of a machine's state. Yet, as discussed under criterion~\ref{crit_Purity}, non-Markovian baths, which retain memories, can revive lost purity.

\subsection{Processing unit or target}
\label{crit_CPU}

A processing unit receives and uses the energy accessed by the machine (criterion~\ref{crit_Energy}). For example, a quantum circuit's processing units are qubits that store quantum information that undergoes logic gates. 
Platforms for gate-based quantum computing include trapped ions~\cite{Bruzewicz_19_Trapped}, ultracold atoms~\cite{Saffman_16_Quantum}, superconducting qubits~\cite{Kjaergaard_20_Superconducting}, photonics~\cite{Slussarenko_19_Photonic}, molecules~\cite{Jones_11_Quantum,Vandersypen_05_NMR}, quantum dots~\cite{Zhang_18_Semiconductor,Zwolak_23_Advances}, color centers in diamond~\cite{Pezzagna_21_Quantum}, and more.
In an autonomous quantum clock (described under criterion~\ref{crit_Time}), a processing unit ticks, emitting excitations.

A machine's purpose can be to operate on a target. For instance, a refrigerator's target is the system that undergoes cooling. An engine's target may be a battery that stores the energy extracted by the engine. A drone's target may be the package to be delivered.

\subsection{Interactions among the machine's components}
\label{crit_Interact}

The interacting components can include the components interfacing with the energy source (criterion~\ref{crit_Energy}),  a processing unit (criterion~\ref{crit_CPU}), a target (criterion~\ref{crit_CPU}), and a timekeeping device (criterion~\ref{crit_Time}). We illustrate interactions with two examples: a quantum absorption refrigerator and a switch.

\subsubsection{Quantum absorption refrigerator}

\begin{figure}[h]
\centering
\begin{subfigure}{0.225\textwidth}
\centering
\includegraphics[width=1\textwidth]{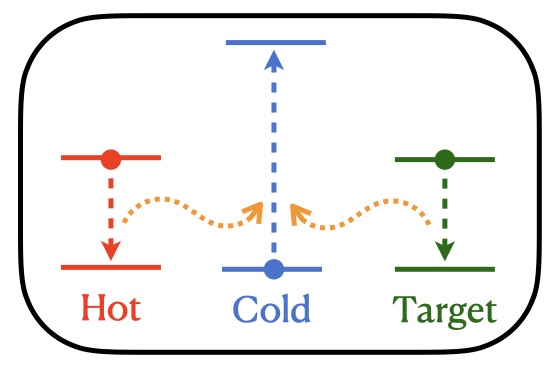}
\caption{\caphead{Quantum absorption refrigerator.}}
\label{fig_Fridge}
\end{subfigure}
\begin{subfigure}{0.225\textwidth}
\centering
\includegraphics[width=1\textwidth]{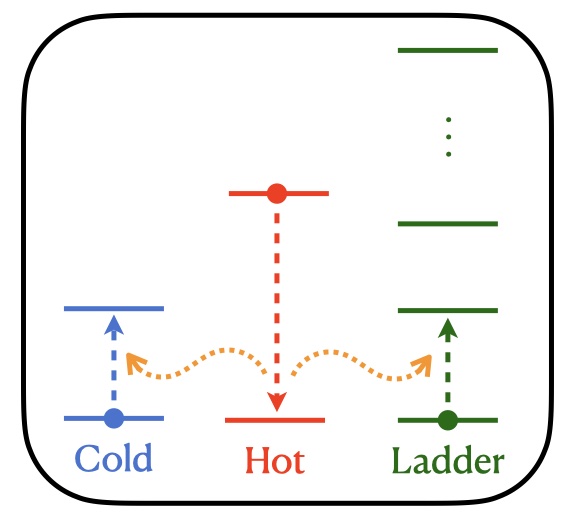}
\caption{\caphead{Autonomous quantum clock.}}
\label{fig_Clock}
\end{subfigure}
\caption{\caphead{Interactions among components:}
The quantum absorption refrigerator and clock evolve under similar Hamiltonians. However, the machines' qudits have different energy gaps: the refrigerator's $\Delta_\hot < \Delta_\cold$, whereas the clock's $\Delta_\hot > \Delta_\cold$. Also, the number of the target's energy gaps can differ between machines. The gaps, with the baths' temperatures ($T_\hot$ and $T_\cold$), determine the direction in which energy flows.}
\label{fig_Interactions}
\end{figure}

A simple quantum absorption refrigerator consists of three qubits: a hot, a cold, and a target qubit (Fig.~\ref{fig_Fridge})~\cite{linden2010,mitchison2019}. 
The hot qubit, $\hot$, evolves under a Hamiltonian $H_\hot = \Delta_\hot  \,  \sigma_z$, wherein $\Delta_\hot > 0$. 
This qubit exchanges heat with a thermal bath at a temperature $T_\hot = 1 / \beta_\hot$. (We set Boltzmann's constant to $\kB = 1$.) Hence $\hot$ begins in the thermal state $e^{- \beta_\hot H_\hot} / Z_\hot$. The partition function is
$Z_\hot  \coloneqq  \Tr ( e^{- \beta_\hot H_\hot} )$. 
The cold qubit, $\cold$, evolves under a Hamiltonian 
$H_\cold = \Delta_\cold \, \sigma_z$ with a larger gap: 
$\Delta_\cold  >  \Delta_\hot$. $\cold$ thermalizes with a bath at the lower temperature $T_\cold  <  T_\hot$, to the state
$e^{-\beta_\cold H_\cold} / Z_\cold$.
The refrigerator cools the target qubit, $\tar$, which evolves under the Hamiltonian
$H_\tar = \Delta_\tar  \,  \sigma_z$, wherein $\Delta_\tar > 0$.

To analyze the absorption refrigerator, we invoke the density operator's statistical interpretation: We can imagine running the refrigerator many times. Each time, $\hot$ begins in an energy eigenstate, $\ket{1}$ or $\ket{0}$, selected according to the Boltzmann distribution $\{ e^{- \beta_\hot \Delta_\hot } / Z_\hot,  \, e^{ \beta_\hot \Delta_\hot } / Z_\hot \}$. The cold qubit begins in an energy eigenstate selected analogously. In an illustrative ``trial,'' $\hot$ begins excited (in $\ket{1}$), while $\cold$ and $\tar$ begin de-excited (in $\ket{0}$). The hot and target qubits emit their excitations into $\cold$, via the three-body interaction
\begin{align}
   \label{eq_Fridge_Inter}
   \ket{1}_\hot  \ket{0}_\cold   \ket{1}_\tar
   \leftrightarrow  \ket{0}_\hot  \ket{1}_\cold   \ket{0}_\tar .
\end{align}
Effective three-body interactions are necessary for autonomous cooling~\cite{linden2010}. They can be effected perturbatively with simultaneous two-body interactions~\cite{hofer2016a,Mitchison_16_Realising,Hewgill_20_Three,Ali_23_Thermally}.
The exchange~\eqref{eq_Fridge_Inter} occurs only if $\Delta_\hot  +  \Delta_\tar  =  \Delta_\cold$---under a condition called strict, or microscopic, energy conservation~\cite{Skrzypczyk_14_Work,Lostaglio_19_Introductory}.
The thermal qubits' gaps and temperatures bias the interaction~\eqref{eq_Fridge_Inter} rightward. Ending in its ground state, $\tar$ has undergone cooling by the refrigerator.

\subsubsection{Switch}

The second interaction example involves a \emph{switch}, a component that keeps time and controls the rest of the machine. Denote by $\switch$ a switch whose Hilbert space has an eigenbasis $\{ \ket{\varphi_j} \}$~\cite[Suppl.~Note~VIII]{Brandao_13_Resource}. If $\switch$ is in state $\ket{\varphi_j}$, then a Hamiltonian $H_j$ evolves the rest of the machine. Under different $H_j$'s, the machine performs different tasks. For example, $H_j$ can denote the $j^\th$ gate available to an autonomous quantum circuit. A Hamiltonian $H_\switch$ evolves the switch's state between $\ket{ \varphi_j }$'s. If the rest of the machine has a constant Hamiltonian $H_\rest$, the total Hamiltonian is
\begin{align}
   \label{eq_Switch_Ham}
   H_\tot
   =  \sum_j  \left(  H_j  \otimes  \ketbra{\varphi_j}{\varphi_j}  \right)
   +  H_\switch  +  H_\rest .
\end{align}
Identity operators $\id$ are implicitly tensored on wherever necessary for each operator to act on the full Hilbert space.

\begin{figure}[hbt]
\centering
\includegraphics[width=.12\textwidth, clip=true]{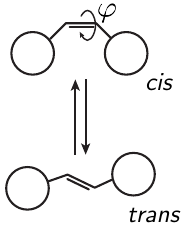}
\caption{\caphead{Photoisomer:} The molecule can switch between \emph{cis} and \emph{trans} configurations as some of its nuclei rotate. Reproduced with permission from Fig.~1 of~\cite{NYH_20_Fundamental}.}
\label{fig_Isomer}
\end{figure}

Photoisomers were argued to contain switches~[App.~D]\cite{NYH_20_Fundamental}.\footnote{ 
Photoisomers are often called molecular switches. Molecular switches should not be confused with the switches described in the previous paragraph---timekeeping switches that control the rest of a machine. Molecular switches contain timekeeping switches, according to~\cite{NYH_20_Fundamental}.}
A photoisomer has two degrees of freedom (DOFs), one nuclear and one electronic. The nuclear DOF, being heavy and slow, determines the electronic DOF's potential landscape. After photoexcitation, some nuclei rotate away from others under $H_\switch = \ell_\varphi^2 / (2 I)$. $\ell_\varphi$ denotes the angular-momentum operator, and $I$ denotes the moment of inertia. $\varphi$ labels the nuclei's relative angular position (Fig.~\ref{fig_Isomer}).
The angles form a continuous set; so the sum in Eq.~\eqref{eq_Switch_Ham} becomes an integral, and $H_j$ becomes $H(\varphi)$.
$H_\switch$ evolves the nuclear configuration from some initial angular position $\ket{\varphi_0}$ to other $\ket{\varphi}$'s. Rotating, the nuclear DOF changes the potential landscape and so the $H(\varphi)$ experienced by the electronic DOF, which forms the rest of the machine.

We have discussed switches under criterion~\ref{crit_Interact} to elucidate their interactions with the rest of their machines. As timekeepers, though, switches belong also under the following criterion.

\subsection{Timekeeping mechanism}
\label{crit_Time}

Timekeeping devices include the switches introduced in the previous subsection; clocks, which tick regularly; and timers, which announce when a programmable amount of time has passed. Below, we review a simple autonomous quantum clock. Then, we discuss timekeeping mechanisms not formed from physical devices. Finally, we delineate three purposes of autonomous quantum machines' timekeeping mechanisms.

\subsubsection{A simple autonomous quantum clock}

If a device is to be autonomous and quantum and to contain a clock, the clock must be autonomous and quantum.
Reference~\cite{Erker_17_Autonomous} introduced a simple, idealized autonomous quantum clock (Fig.~\ref{fig_Clock}). The clock contains a hot qubit $\hot$ and a cold qubit $\cold$, like the absorption refrigerator under criterion~\ref{crit_Interact}. $\hot$ has the larger gap here, however: 
$\Delta_\hot > \Delta_\cold$. The clock contains also a ladder $\ladder$ of $d$ energy levels: $H_\ladder = \sum_{j=0}^{d-1}  j \Delta  \,  \ketbra{j}{j}$. 
$\ladder$ begins in its ground state, $\ket{0}$. 
The qudits (multilevel quantum systems) undergo a three-body interaction similar to the quantum absorption refrigerator's Eq.~\eqref{eq_Fridge_Inter}: 
\begin{align}
   \label{eq_Clock}
   \ket{1}_\hot  \ket{0}_\cold  \ket{0}_\ladder  
   \leftrightarrow  \ket{0}_\hot  \ket{1}_\cold  \ket{j}_\ladder .
\end{align}
The ladder state $\ket{j}$ satisfies strict/microscopic energy conservation through
$\Delta_\hot  =  \Delta_\cold  +  j \Delta$.
The condition $\Delta_\hot > \Delta_\cold$ biases the interaction~\eqref{eq_Clock} rightward.
The hot qubit, de-exciting (undergoing $\ket{1}  \mapsto  \ket{0}$), drives the cold qubit upward in energy (through $\ket{0}  \mapsto  \ket{1}$) and drives the ladder system upward 
(through $\ket{0}  \mapsto  \ket{j}$). The baths reset $\hot$ and $\cold$, and the process repeats. Upon reaching its top rung, the ladder system ticks---emits an excitation---returning to $\ket{0}$.

The foregoing model is a simple one intended to capture the basic physics. Such a clock would keep time poorly. Schwarzans \emph{et al.} mitigate this challenge with a more complex clockwork~\cite{Schwarzhans_21_Autonomous}: Each ladder transition interacts with multiple pairs of hot and cold qubits. As the number of pairs grows and as $d$ grows, ``a perfect clockwork can be approximated arbitrarily well''~\cite{Schwarzhans_21_Autonomous}. More precisely, denote by $\rho_t$ the ladder's time-$t$ state. When a tick should happen, all the probability weight occupies the top rung: $\bra{d-1} \rho_t \ket{d-1} = 1$. Otherwise, no probability weight does: $\bra{d-1} \rho_t \ket{d-1} = 0$. Granted, entropy production accompanies timekeeping; a tick requires not only probability concentration, but also emission to the environment. This entropy production decrees a tradeoff between accuracy and resolution~\cite{Schwarzhans_21_Autonomous,Meier_23_Fundamental}. Conveniently, autonomous quantum clocks appear to be well-positioned to operate near the fundamental bounds on quality~\cite{Schwarzhans_21_Autonomous}. 
More work is required to advance the theory and realization of autonomous quantum clocks to such a level, however, as discussed in Sec.~\ref{sec_Outlook}.

Possible physical realizations include cavity quantum electrodynamics (QED). 
Atomic energy levels may realize the ladder. So may transmon superconducting circuits, whose lowest $d$ energy levels can serve as $d$-level qudits~\cite{You_11_Atomic,Morvan_20_Qutrit,Blok_20_Quantum}.
Ticking, the ladder would emit a photon by spontaneous emission into the cavity.
Challenges include preventing the ladder from ticking until it reaches $\ket{d}$.
Another challenge is detecting single photons: Superconducting qubits emit microwave-frequency photons, whose low energies can escape detection. However, single-microwave-photon detectors formed from superconductors~\cite{Inomata_16_Single,DElia_23_Stepping} and calorimeters~\cite{Roukes_99_Yoctocalorimetry,Karimi_20_Reaching,Kokkoniemi_20_Bolometer,Lee_20_Graphene,Katti_22_Josephson,Pekola_22_Ultrasensitive} are under development.

Figures of this clock's merit include its \emph{accuracy}, $N$~\cite{Schwarzhans_21_Autonomous}. 
Denote by $\bar{t}$ the average time interval between ticks; and by $\Delta t$, the standard deviation in that time interval.
The accuracy is $N \coloneqq (\bar{t} / \Delta t)^2$.
To interpret it, we consider the limit as the number of ticks, assumed to be distributed independently and identically, approaches infinity.
$N$ equals the number of ticks that pass, on average, before the clock is off by one tick.
One can improve $N$ and other figures of merit by complicating the clockwork~\cite{Schwarzhans_21_Autonomous}. We discuss figures of merit further under criterion~\ref{crit_Output}.

\subsubsection{Timekeeping mechanisms other than physical devices}

Not all autonomous quantum machines need timekeeping devices as physical components. The finiteness of a machine's energy resources (criterion~\ref{crit_Energy}) can induce a timer, as can the machine's coherence, we explain under criterion~\ref{crit_Off}. Furthermore, some autonomous quantum engines operate continuously: They undergo fixed dynamics, rather than cycles formed from discrete (timed) strokes~\cite{Kosloff_14_Quantum}. 

\subsubsection{Timekeeping mechanisms' purposes}

An autonomous quantum machine's timekeeper fulfills three purposes.
First, the timekeeper ensures that the machine initiates an action at the right time. For example, a quantum circuit should begin each gate during the appropriate part of a computation. A Rydberg-atom computer~\cite{Saffman_16_Quantum} might autonomously perform a Rydberg-blockade entangling gate~\cite{Jaksch_00_Fast} as follows. 
Let the circuit have an autonomous quantum clock that ticks by emitting an excitation. Two excitations can boost two atoms to their Rydberg (high-energy) states. The excited atoms will repel each other, entangling. Two more excitations may stimulate emissions from the atoms, which will return to their ground states.

An implementation of autonomous two-qubit Rydberg-atom gates is being designed~\cite{JAMG_23_Three}; we sketch the idea here.
To manipulate Rydberg atoms, the clock ticks will need to satisfy stringent requirements---to qualify as pulses of high intensities, specific durations, etc. These requirements may be satisfied by a passive mode-locked laser~\cite{Silfvast_04_Laser,Innerhofer_03_60}. Such a laser contains a medium that absorbs light whose intensity lies below a certain threshold. Once the intensity exceeds the threshold, the laser emits a pulse. Mode-locking techniques stabilize and regularize the laser pulses, albeit at the expense of pulse duration~\cite{Silfvast_04_Laser}. 
Example techniques include colliding-pulse and additive-pulse mode locking. The laser will emit pulses until depleting its atoms of energy. At this point, the Rydberg-atom qubits will quit undergoing gates.

One might be concerned that an autonomous quantum circuit cannot undergo single-qubit gates: classical external fields often effect single-qubit rotations. However, autonomous quantum circuits may implement Brownian circuits~\cite{Brown_12_Scrambling,Lashkari_13_Towards}. Brownian circuits have recently elucidated properties of chaos, randomness, and scrambling~\cite{Shenker_15_Stringy,Zhou_19_Operator,Bentsen_21_Measurement}. Such circuits can consist solely of two-qubit (and even nearest-neighbor) interactions and so may amenable to autonomous quantum computation. 
   
Second, the timekeeper ensures that the machine performs an action for the desired amount of time. For example, a quantum circuit implements a gate by effecting some Hamiltonian for the correct time interval. Xuereb \emph{et al.} calculated the average fidelity $\bar{\mathcal{F}}$ of a CNOT gate $U$ approximated with an autonomous quantum clock of accuracy $N$~\cite{Xuereb_23_Impact}. 
The evolution implemented is a channel $\mathcal{E}$. Denote by $d \psi$ the Haar measure (loosely speaking, the uniform measure) over the set of two-qubit pure states $\ket{\psi}$. The average fidelity is defined as
$\bar{\mathcal{F}} 
\coloneqq \int  d \psi 
 \langle \psi | U^\dag  \,  \mathcal{E} ( \ketbra{\psi}{\psi} )  \,  U  \ket{\psi}$.
 According to~\cite{Xuereb_23_Impact},
$\bar{\mathcal{F}} = (2 + e^{-\pi^2 / (2N) } ) / 3$. 
The clock accuracy $N$ exponentially influences the correction to the constant $2/3$.
Third, a timekeeper ensures that the machine turns off, satisfying criterion~\ref{crit_Off}, upon completing its task. 

%
\subsection{Structural integrity}
\label{crit_Structure}

Imagine placing three atoms beside each other, as with optical tweezers. If left unattended, the atoms will drift apart. A machine's components must not separate, lest they cease to satisfy the interaction criterion~\ref{crit_Interact}. Denote by $r$ the distance between two components. 
If they carry electric charges, the Coulombic interaction potential between them weakens as $r^{-1}$. Alternatively, the components may be atoms or molecules excitable to Rydberg states. If so, the Rydberg-blockade potential decreases as $r^{-6}$~\cite{Saffman_16_Quantum}. 
An autonomous quantum machine formed from multiple atoms, ions, etc. requires trapping, as by lasers. Laser light, forming a wave, provides a time-dependent external potential. Hence a laser-trapped machine can be autonomous only if the trapped spatial DOFs play no role in the machine's functioning. Those DOFs' Hamiltonian must commute with the microscopic Hamiltonian that governs the machine's operation.

Yet a machine need not consist of physically separated components. A photoisomer functions as an autonomous quantum energy transmitter, as explained in the introduction. The molecule contains an autonomous quantum clock, according to~\cite{NYH_20_Fundamental}. Similarly, photosynthetic complexes~\cite{Li_23_Single} and enzymes~\cite{Klinman_13_Hydrogen} are autonomous machines modeled usefully with quantum theory. A molecule can therefore form an autonomous quantum machine. So can an atom, arguably.\footnote{
The atomic levels analyzed in~\cite{Scovil_59_Three,Geusic_67_Quantum,linden2010} interact with heat baths whose frequencies are filtered. One might count the filters as parts of the machine.}
For example, three atomic energy levels can form an autonomous engine or refrigerator~\cite{Scovil_59_Three,Geusic_67_Quantum,linden2010}. 
Not only natural particles, but also artificial devices can form self-contained autonomous quantum machines:
Superconducting qubits cannot separate if printed on the same chip. Neither can the dopants that host qubits on solid-state surfaces; and neither can the gate electrodes that define quantum dots, being grown or printed on a semiconductor chip.

%
\subsection{Sufficient purity}
\label{crit_Purity}

In this subsection, we define purity and compare it with coherence. Next, we explain the \emph{sufficient} in the subsection's title. We then discuss the tradeoff between purity and accessible energy (criterion~\ref{crit_Energy}). Non-Markovianity may soften the tradeoff.

Purity is defined as follows. Denote by $\rho$ an arbitrary quantum state (density operator) defined on a $\Dim$-dimensional Hilbert space $\Hil$. $\rho$ has an amount $\pure(\rho) \coloneqq \Tr (\rho^2)$ of purity. If $\rho$ is pure (if $\rho = \ketbra{\psi}{\psi}$ for some $\ket{\psi} \in \Hil$), then $\pure(\rho) = 1$. If $\rho$ is the maximally mixed state $\id / \Dim$, then $\pure(\rho) = 1 / \Dim$. A machine's processing unit, target, timekeeping device, and/or output might require purity at various times.

One might expect our criteria to include coherence, rather than purity. In quantum thermodynamics, coherence relative to the energy eigenbasis often serves as a resource (e.g.,~\cite{Vaccaro_08_Tradeoff,Aaberg_14_Catalytic,Lostaglio_15_Description,Cwiklinski_15_Limitations,Korzekwa_16_Extraction,Marvian_16_How,Marvian_16_From,Kwon_18_Clock,Woods_19_Autonomous,NYH_20_Fundamental,Marvian_20_Coherence}). 
To quantify this coherence, we suppose that $\rho$ evolves under a Hamiltonian 
$H = \sum_j E_j \ketbra{j}{j}$.
For $j \neq k$, the off-diagonal element $\rho_{jk} \coloneqq \bra{j} \rho \ket{k}$ is a coherence. Mode $\omega$ of coherence is defined as
$\rho^{(\omega)}  
\coloneqq  \sum_{j, k \, : \, E_j - E_k = \omega}
\rho_{jk}  \ketbra{j}{k}$
and quantified with, e.g., 
$\sum_{j, k : E_j - E_k = \omega}  | \rho_{jk} |$~\cite{Marvian_14_Modes,Lostaglio_15_Quantum}.
Despite its applications in timekeeping and work extraction, such coherence can be undesirable. For instance, qubits are often initialized to near their ground states~\cite{Taranto_23_Landauer}, $\ket{0}$, before a quantum computation. Resetting computational qubits, an autonomous quantum refrigerator destroys coherences relative to the energy eigenbasis. Still, the refrigerator enhances the qubits' purities. 
Hence our criteria's including purity, rather than coherence. 

Our criteria include \emph{sufficient} purity for two reasons: (i) Different machine components may require different amounts of purity. (ii) One component may require different amounts of purity at different times. We illustrate with the quantum absorption refrigerator described under criterion~\ref{crit_Interact}. The hot and cold qubits, interacting with finite-temperature baths, are always mixed. The target requires purity---closeness to $\ket{0}$---but only when the protocol ends. In contrast, suppose that an autonomous quantum clock's ladder (Fig.~\ref{fig_Clock}) is mixed at any time. The clock's accuracy seems likely to suffer~\cite{Erker_17_Autonomous}.
Similarly, a quantum circuit must retain enough purity to meet the threshold for fault-tolerant quantum error correction throughout its computation~\cite{Knill_98_Resilient,Kitaev_03_Fault,Aharonov_08_Fault}. 
The error rate is 0.6\%--1\% for the two-dimensional surface code~\cite{Raussendorf_07_Fault,Fowler_09_High}.
Via external control, two-qubit fidelities of $\geq 99.5\%$ have been achieved with trapped ions~\cite{Bruzewicz_19_Trapped},
superconducting qubits~\cite{Kjaergaard_20_Superconducting}, and
neutral atoms~\cite{Evered_23_High}.

The sufficient-purity criterion trades off with the accessible-energy criterion~\ref{crit_Energy}. As described below~\ref{crit_Energy}, an autonomous quantum machine may extract energy from heat baths. The baths reduce the machine's $\pure(\rho)$. 
However, non-Markovianity can revive a machine's purity~\cite{Breur_16_Colloquium}.
A non-Markovian bath retains a memory; information, upon entering the bath from the machine, can recollect and act back on the machine. Non-Markovianity can arise from strong system--bath couplings, small baths, low temperatures, and initial system--environment couplings. One can engineer non-Markovianity from a Markovian environment: One would mediate machine--environment interactions through a memory-retaining interface~\cite{Li_10_Non}. Mediating superconducting qubits, the interface could be a resonator or a cavity. A long lifetime or hysteresis would provide memory.
Appendix~\ref{app_NonMarkov} illustrates non-Markovianity's potential for reviving a machine's purity. The model there can be realized with cavity QED.
Non-Markovianity, the appendix shows, can help baths achieve the energy criterion~\ref{crit_Energy} while endangering the purity criterion~\ref{crit_Purity} less than Markovian baths do.

\subsection{Output worth the input}
\label{crit_Output}

A machine's output is intended to fulfill the machine's purpose. How effectively the output fulfills the purpose depends on figures of merit. Any agent running a machine can choose their favorite figures of merit, as well as their thresholds for acceptable figure-of-merit values. Due to this analysis's agent-centric nature, no Perspective can prescribe ``one figure of merit to rule them all,'' even for one machine. Instead, we illustrate with eight figures of merit for four machines. Then, we detail input costs, including energy, time, and control.

\subsubsection{Example figures of merit}

First, a clock outputs ticks (emitted excitations). Its figures of merit include the accuracy, 
$N  \coloneqq ( \bar{t} / \Delta t )^2$, described under criterion~\ref{crit_Time}. Another figure of merit is the resolution~\cite{Erker_17_Autonomous}. 
Recall that $\bar{t}$ denotes the average time between successive ticks. 
The resolution is $1 / \bar{t}$. 

Second, a quantum circuit outputs a state $\sigma$ that approximates an ideal $\rho$. Figures of merit can quantify the distance between the states. For example, the fidelity is 
$\big( \Tr \sqrt{ \sqrt{ \sigma }  \,  \rho \,  \sqrt{ \sigma }  }  \big)^2$.

Third, an engine outputs work. Denote by $\dot{W}$ the power and by $\dot{Q}_\hot$ the current of heat flowing from the engine's hot bath. Figures of merit include the steady-state efficiency, $\eta = \dot{W} / \dot{Q}_\hot$; the power, $\dot{W}$; and the efficiency at maximum power, 
$\max_{\eta}  \{ \dot{W} \}$. 

Fourth, a refrigerator outputs a cooled target, $\tar$, whose final temperature quantifies the refrigerator's effectiveness. So does the refrigerator's steady-state coefficient of performance (COP), the analogue of the engine's efficiency:
Denote by $\dot{Q}_\tar$ the current of heat extracted from $\tar$ and by $\dot{Q}_\hot$ the current of heat flowing from the hot bath. The steady-state COP is $\dot{Q}_\tar / \dot{Q}_\hot$~\cite{mitchison2019}. 

%
\subsubsection{Input costs}

Each above figure of merit measures an output's quality. The efficiency and COP compare outputs with their input costs. Comprehensive figures of merit capture such comparisons~\cite{Auffeves_22_Quantum}. Inputs can include energy, time, and control. We illustrate energy with a single-atom engine and a quantum refrigerator. Then, we distinguish two input costs: the fabrication of an autonomous quantum machine and the preparation of a fabricated machine for one trial. If the preparation is simple enough, and the machine executes enough trials, the total output can outweigh a large fabrication cost.

As explained in the introduction, one can expect $\approx 1$ eV per cycle from a natural-atom engine. Calculating a heat engine's efficiency, one counts as input only the heat absorbed by the engine from its hot environment. Yet cooling and initializing the engine can cost orders of magnitude more than $\approx 1$ eV. The quantum engines realized so far, to our knowledge, cost more work than they output. Once experimentalists finish exploring the fundamentals of quantum engines, quantum engines will merit realization only if they meet the present criterion---only if their outputs merits the total input. 

Reference~\cite{Ali_23_Thermally} showed how an autonomous superconducting-qubit refrigerator can cost little input. The quantum refrigerator sits inside a dilution refrigerator, which is already cold because it hosts a superconducting-qubit quantum computer. Keeping the quantum refrigerator cold (at a temperature low enough to support quantum phenomena) therefore costs negligible energy per qubit.
The dilution refrigerator consists of layers, which progress from hot to cold when traversed from outermost to innermost. The innermost layer can serve as the quantum refrigerator's cold bath. An outer layer can serve as the hot bath, connected to the quantum refrigerator via a waveguide. Operating the quantum refrigerator therefore costs little beyond the energy sunk into the quantum computer. Other autonomous quantum machines might achieve a high output--input ratio similarly if slotted into appropriate environments.

The autonomous quantum refrigerator illustrates how reusable machines cost two input processes: fabrication, as well as preparations for individual trials. Fabrication is the creation of the machine. Fabricating an autonomous quantum refrigerator involves creating a nanoscale chip. Similarly, fabricating a classical drone can cost a factory---an enormous amount of money, person hours, real estate, and equipment. Once a factory exists, though, creating and programming drones is relatively simple. The factory can easily produce many drones, each of which may easily be programmed to deliver many packages. The total number of packages delivered can outweigh the large fabrication cost, plus the smaller single-delivery cost times the number of deliveries. Similarly, once one fabricates a quantum refrigerator, initiating it to cool a target is simple: the available hot environment prepares one qubit in a high-temperature thermal state, and the cold environment prepares another qubit in a low-temperature state~\cite{Ali_23_Thermally}. Additionally, upon fabricating one quantum-refrigerator chip, a lab or company may fabricate others easily.  The small single-trial cost, times the total number of trials, should outweigh the fabrication.

\subsection{Ability to switch off after completing assignment}
\label{crit_Off}

At least three mechanisms can spur an autonomous quantum machine to shut down. First, consider a machine that contains an autonomous quantum clock (criterion~\ref{crit_Time}). The clock can galvanize not only steps in the rest of the machine's operation (e.g., gates implemented at the right times), but also a halt. Second, the machine accesses only a finite amount of energy (criterion~\ref{crit_Energy})---for instance, finite-size hot and cold baths. Depleting the energy source winds the machine down. Consider a machine formed from natural or artificial atoms in a cavity. The cavity could be populated initially with finitely many photons, which would provide a finite energy source.

Third, components of the machine can require varying degrees of purity to operate (criterion~\ref{crit_Purity}). Hence the components' coherence times limit the operation time. We illustrate with a qubit with a ground state $\ket{0}$, excited state $\ket{1}$, and time-evolving density operator $\rho(t) = \sum_{j,k = 0}^1  \rho_{jk}(t)  \ketbra{j}{k}$, for $t \geq 0$~\cite{Preskill_15_Physics}.
Over the amplitude-damping time $T_1$, the excited-state weight $\rho_{11}(t)$ decays to $1/e$ of its initial value: $\rho_{11}(T_1) = \rho_{11}(0) / e$.
Over the phase-damping time $T_2 < T_1$, the off-diagonal elements decay similarly:
$\rho_{jk} (T_2)  =  \rho_{jk}(0) / e$ $\forall j \neq k$.
Of the quantum-computing platforms, nuclear spins excel at maintaining coherence: Europium dopants in a solid (yttrium orthosilicate) have achieved a six-hour $T_2$ time~\cite{Zhong_15_Optically}.
Isolated $^{171}$Yb$^+$ ions have exhibited $T_2$ times of over an hour~\cite{Wang_21_Single},
although collections of ions decohere more quickly~\cite{Bruzewicz_19_Trapped}.

%
\subsection{Mobility (optional)}
\label{crit_Mobility}

DiVincenzo listed two optional criteria necessary for transmitting information from place to place. Our optional criteria,~\ref{crit_Mobility} and~\ref{crit_Interop}, concern the transmission of machines and of messages between machines. Mobility would benefit autonomous quantum machines including delivery drones, sensors that navigate to their targets, and machines that build molecules (or other machines). External potentials and accessible energy (criterion~\ref{crit_Energy}) can help machines achieve directionality.

We illustrate with autonomous classical nanowalkers, which may inspire builders of autonomous quantum walkers. Reference~\cite{Yin_04_Unidirectional} reports on a nanowalker, formed from a DNA strip, walking along a track consisting of more DNA strips. The nanowalker burns ATP as fuel. Enzymes ensure the nanowalker's directionality: One enzyme ligates (joins together) the walker and the next site, and another enzyme cleaves the walker from the previous site.

Intriguingly, single organic molecules may serve as quantum sensors: Pentacene in a $p$-terphenyl crystal has spin properties sensitive to its environment. These spin properties have been read out optically~\cite{Wrachtrup_93_Optical,Wrachtrup_93_Optically}. 
Whether organic-molecule detectors can merge with organic-molecule nanowalkers is far from clear, however. For starters, the detectors require certain absorption and stability properties; not any organic molecule will serve.
Nonetheless, in the blue-sky spirit of this Perspective, one might imagine quantum sensors that propel themselves to their targets.

\subsection{Interoperability (optional)}
\label{crit_Interop}

Autonomous quantum machines may communicate with each other and work together. Communications may occur via (at least) two mechanisms.

First, machine $\A$ may emit a signal for machine $\B$ to absorb. For example, we envisioned an autonomous quantum clock emitting a photon absorbed by an autonomous Rydberg-atom circuit (under criterion~\ref{crit_Time}).\footnote{
Under criterion~\ref{crit_Time}, we cast the clock as a part of the circuit. Here, we cast the clock as separate but as aiding the circuit.}
$\A$ and $\B$ must satisfy three requirements: 
\begin{enumerate}[label=(\arabic*)]

   \item  \label{item_Detect}
   $\B$ must be physically able to detect $\A$'s signal. For example, denote by $\hbar \omega$ the excitation's energy, by $\Omega_\B$ the bandwidth of $\B$, and by $E_0$ the center of $\B$'s energy spectrum. The energy must lie in the bandwidth: 
   $\hbar \omega  \in  [E_0  -  \Omega_\B/2,  \,  E_0  +  \Omega_\B/2]$. 
   
   \item  \label{item_Prob}
   The signal must, given condition~\ref{item_Detect}, have a sufficiently high probability of affecting $\B$. The excitation's momentum should direct the signal toward $\B$, and no intervening medium should swallow the signal. Once the excitation arrives, $\B$ should have a high probability of absorbing it. Fermi's golden rule can mediate this probability, governing the rate $\Gamma_{\text{i} \mapsto \text{f}}$ at which $\B$ jumps from a state $\ket{\rm i}$ to any energy-$E_{\rm f}$ state $\ket{\rm f}$:
$\Gamma_{\text{i} \mapsto \text{f}} 
=  \frac{2 \pi}{\hbar}  \,  
| \bra{ \text{f} }  H'  \ket{ \text{i} } |^2  \,
\mu( E_{\rm f} )$.
$H'$ denotes the excitation-induced perturbation to $\B$'s Hamiltonian. $\mu(E)$ denotes the density of $\B$'s energy-$E$ states.
If $\B$ absorbs the signal through a photodetector, the external efficiency quantifies how effectively $\B$ satisfies requirement~\ref{item_Prob}. One can achieve a high efficiency upon matching $\A$'s and $\B$'s impedances. Impedance matching has enabled nearly perfect photon absorption by a cavity-QED detector~\cite{Koshino_13_Implementation,Inomata_14_Microwave}. The scheme leveraged the detector's $\Lambda$-type energy-level structure, albeit nonautonomously. Apart from impedance matching, realistic interconnects entail other engineering challenges, such as losses in the interconnects.

   \item After absorbing a signal, $\B$ might take time to reset before being able to absorb another signal. Denote this \emph{dead time} by $\tau_\dead$. Suppose that $\A$ sends multiple signals. The time interval $\tau$ between them should be 
   $\tau  \geq  \tau_\dead$. Photodetectors have dead times, 
   a typical example of which is 20 ns (e.g.,~\cite{Tomm_20_Bright}).
   Also, consider a photoisomer that has absorbed a photon and rotated into a metastable configuration. The molecule must re-equilibrate before undergoing another photoexcitation. The metastable configuration can have a half-life of two days, in solution at room temperature, if the photoisomer is azobenzene~\cite{Bandara_12_Photoisomerization}.
   
\end{enumerate}

Under the second signaling mechanism, machine $\A$ arrives near $\B$, changing the external potential experienced by $\B$. Again, we illustrate with a photoisomer. Some of its nuclei rotate through many $\ket{\varphi}$'s, in the notation under criterion~\ref{crit_Interact}. Advancing so, the nuclei change the electronic DOF's potential landscape---change $H(\varphi)$. 

In another example, a qubit and a bosonic mode undergo a dispersive interaction. 
Such interactions are common in cavity QED, the qubit manifesting as a natural or artificial atom. The dispersive interaction is effective (approximate), and we omit the derivation~\cite{Blais_21_Circuit}. 
During it, one attributes to the qubit an effective gap $2 \Delta$; to the mode an effective frequency $\omega$, a creation operator $a^\dag$, and an annihilation operator $a$; and to the coupling a strength $\chi$. We set $\hbar = 1$. The dispersive Hamiltonian is
\begin{align}
   \Delta  \,  \sigma_z  +  \omega  a^\dag a  +  \chi  \sigma_z  a^\dag a .
\end{align}
If the qubit is excited (in $\ket{1}$), it effectively adds $\chi$ to the mode's frequency. Analogously, if the mode is occupied, it effectively adds $2 \chi$ to the qubit's gap. Each subsystem therefore affects the other's Hamiltonian. 

%
%
%
\section{Outlook}
\label{sec_Outlook}

We have proposed DiVincenzo-like criteria for autonomous quantum machines. Eight criteria, we regard as necessary for most autonomous machines' useful operation. Satisfying the remaining two criteria, machines can move and interface. 
Appendix~\ref{sec_Exclusions} concerns two topics adjacent to our criteria: instructions (which emerge from other criteria) and measurements (which not all useful autonomous quantum machines require.)

We emphasize the \emph{useful} in this section's second sentence: The literature has demonstrated that autonomous quantum machines can be designed and, with effort, realized experimentally. Autonomous quantum machines should now progress from curiosities to tools, like their classical counterparts. 
We hope that our criteria guide the progression, along with our observations about platforms' abilities to realize these criteria. Many platforms offer promise: superconducting qubits, molecules, neutral atoms, trapped ions, quantum dots, single-electron boxes, and thermoelectric systems, as well as carbon nanotubes~\cite{Parrondo_23_Information}.

We now present five challenges in building autonomous quantum machines, as well as possible solutions. Afterward, we discuss a yet-more ambitious goal for a future Perspective: autonomous quantum machines that outperform all classical counterparts.
\begin{enumerate} 

   \item \label{item_Enviro}
   The community must identify more settings that (i) contain energy sources usable by autonomous quantum machines (criterion~\ref{crit_Energy}) and (ii) support autonomous machines' quantum behaviors without costing users significant extra work (criterion~\ref{crit_Output}). We have identified one setting: a dilution refrigerator that has already been cooled to support quantum computation~\cite{Ali_23_Thermally}. Biochemistry may furnish other settings. Granted, biochemical systems are warm and wet, tending to supprehence purity. Yet biochemistry supports quantum behaviors by photoisomers, enzymes, and more, as discussed in the introduction. Furthermore, biochemical systems are far from equilibrium and so contain free energy.
   
   \item Usable energy---especially energy extracted from heat baths---trades off with purity, as discussed below criterion~\ref{crit_Purity}. Four strategies suggest themselves: (i) Take advantage of the ``sufficient'' in the ``sufficient purity'' criterion: Identify when purity is necessary, and waste no effort on maintaining unnecessary purity. (ii) Situate autonomous quantum machines in environments that stabilize desirable quantum states~\cite{BohrBrask_15_Autonomous}, using engineered dissipation~\cite{Harringon_22_Engineered}. The \emph{engineered} may suggest that maintaining the environment requires work from an agent. This impression is misleading, however. For example, ``the Zeno effect [ \ldots ] is largely passive'' and can raise a system's probability of remaining in its ground state~\cite{Harringon_22_Engineered}. Engineered dissipation generalizes non-Markovian environments, discussed in App.~\ref{app_NonMarkov}.
   (iii) Resolving challenge~\ref{item_Enviro} can resolve this tradeoff challenge.
   
   \item \label{item_Motivate_auto}
   Realizing controlled autonomous quantum machines is difficult. For example, companies, governments, and universities have already poured investments into controlled quantum computers, whose construction will require many more years. Realizing autonomous quantum machines requires advances beyond controlled quantum machines. Hence experimentalists may lack the motivation to introduce autonomy. We identify three mitigating factors.
   
   First, a practical approach to autonomy runs through partial autonomy. One might remove external control from a quantum machine step by step. For example, a quantum computer undergoes a preparation procedure, a transformation, and a measurement~\cite{NielsenC10}. One can grant the quantum computer partial autonomy by implementing the preparation procedure~\cite{Ali_23_Thermally} or gates~\cite{JAMG_23_Three} autonomously. Measurement appears to require action by a classical system and so not to be implementable by an autonomous quantum machine. Therefore, \emph{autonomous quantum computer} appears to be an oxymoron. However, autonomous quantum state preparations and autonomous quantum circuits can still be useful.
   
   Second, (partial or complete) autonomy may assist the quantum machines being built now. For instance, autonomous quantum refrigerators can reset qubits in quantum computers~\cite{JAMG_23_Three}. Also, as discussed in the introduction, pruning control wires may improve superconducting-qubit quantum circuits' scalability and coherence times. 
   
   Third, autonomous quantum machines can enable fundamental discoveries, in addition to performing tasks autonomously. For example, autonomous quantum clocks may enable us to test fundamental limits on timekeeping---tradeoffs involving dissipation and accuracy~\cite{Meier_24_Precision}.
   
   \item \label{item_Clock_exp}
   Autonomous quantum clocks may feature in other autonomous quantum machines, such as autonomous quantum circuits (criterion~\ref{crit_Time}). However, autonomous quantum clocks have yet to be realized experimentally. Progress on this challenge is already underway. An autonomous classical clock was recently realized experimentally. Furthermore, it was presented as paving a path toward a quantum analogue. So has the autonomous-quantum-refrigerator experiment~\cite{Ali_23_Thermally} paved a path: reversing the refrigerator and altering its parameters enables a simple clock~\cite{Erker_17_Autonomous}, as discussed under criterion~\ref{crit_Time}. Once the simple clock is realized, increasing the clockwork's complexity can improve the accuracy and resolution~\cite{Schwarzhans_21_Autonomous}.

   \item The theory of autonomous quantum clocks has remained abstract. Idealizations must be identified and removed. The theoretical quantum-thermodynamics community has already begun to address this challenge: Ref.~\cite{Schwarzhans_21_Autonomous} introduced a complex clockwork that improves the simplest autonomous quantum clock's accuracy and resolution~\cite{Erker_17_Autonomous}. Other opportunities for enhanced modeling include the ladder's coupling to the environment. The ladder has been assumed to tick only after most of its probability weight has reached the top rung. In reality, lower rungs, too, can couple to the environment. Such realities will naturally be addressed as autonomous quantum clocks are built experimentally---as challenge~\ref{item_Clock_exp} is addressed.

\end{enumerate}

Overcoming the above five challenges is necessary for realizing diverse useful autonomous quantum machines. A sixth challenge is unnecessary for achieving that goal but merits mentioning: developing diverse autonomous quantum machines that outperform all classical counterparts. We have not discussed that goal because building useful autonomous quantum machines is sufficiently difficult and worthwhile. Once many such machines exist, however, beyond-classical performance will constitute a natural next step. Granted, autonomous quantum circuits must outperform all classical counterparts to be useful. However, one example does not justify the inclusion of beyond-classical performance in our general criteria. Furthermore, criterion~\ref{crit_Output} (output worth the input) can encode beyond-classical performance when tailored to quantum circuits.

%
%
\begin{acknowledgments}
N.Y.H. thanks Kenneth~Brown for input about castle-in-the-air autonomous quantum machines, David~Limmer for input about chemical machines, Noah Lupu-Gladstein for input about photodetectors, Misha Lukin for input about Rydberg-atom operations, John~Preskill for input about superconducting-qubit chips, and Nathan Schine for input about quantum-computing platforms.
N.Y.H. and J.A.M.G. acknowledge support from the National Science Foundation 
(QLCI grant OMA-2120757), 
the John Templeton Foundation (award no. 62422),
and NIST grant 70NANB21H055\_0. 
P.E. and M.H. acknowledge support from the European Research Council (Consolidator grant ``Co-coquest'' 101043705),
the European flagship on quantum technologies (``ASPECTS'' consortium 101080167), and
FQXi (FQXi-IAF19-03-S2, within the project ``Fueling quantum field machines with information'').
S.G. acknowledges support from the Knut and Alice Wallenberg foundation via the Wallenberg Centre for Quantum Technology (WACQT), from the European Research Council (Grant 101041744 ESQuAT), from the European Union (Grant 101080167 ASPECTS), and from the Swedish Research Council (Grant 2021-05624).
\end{acknowledgments}

\begin{appendices}

\onecolumngrid

\renewcommand{\thesection}{\Alph{section}}
\renewcommand{\thesubsection}{\Alph{section} \arabic{subsection}}
\renewcommand{\thesubsubsection}{\Alph{section} \arabic{subsection} \roman{subsubsection}}

\makeatletter\@addtoreset{equation}{section}
\def\theequation{\thesection\arabic{equation}}

\section{Non-Markovianity's potential for reviving a machine's purity}
\label{app_NonMarkov}


We follow~\cite{Breur_16_Colloquium}. That reference presents a setup common in cavity quantum electrodynamics: Consider a qubit $\Sys$ governed by a Hamiltonian $H_\Sys = \Delta \, \sigma_z$ with an energy splitting $2 \Delta > 0$. 
As usual, we label the ground state as $\ket{0}$ and the excited state as $\ket{1}$.
A bosonic environment $\Env$ evolves under a Hamiltonian 
$H_\Env = \sum_\ell  \omega_\ell  \,  a_\ell^\dag a_\ell$. 
Mode $\ell$ corresponds to energy $\omega_\ell$, to an annihilation operator $a_\ell$, and to a creation operator $a_\ell^\dag$. 
The mode begins in its vacuum state.
In the rotating-wave approximation, $\Sys$ couples to $\Env$ via the Jaynes--Cummings Hamiltonian
$H_\inter  =  \sum_\ell \big(  
g_\ell  \,  \sigma_+  a_\ell  +  g_\ell^*  \sigma_-  a_\ell^\dag  \big)$.
The qubit has raising and lowering operators 
$\sigma_+  \coloneqq  \ketbra{1}{0}$  and  $\sigma_-  \coloneqq  \ketbra{0}{1}$,
and the $g_\ell$'s denote coupling constants. 
Denote by $\rho_{\Sys/\Env}(t)$ the time-$t$ reduced state of $\Sys/\Env$ in the interaction picture. Let $\Sys$ and $\Env$ begin in a product state 
$\rho_\Sys(0) \otimes \rho_\Env(0)$,
$\rho_\Env(0)$ being the bath's vacuum state.

$\Sys$ evolves under a linear, completely positive
%
map $\Phi_t$ as
$\rho_\Sys(t)  =  \Phi_t   \,  \rho_\Sys(0)$.
The time-evolved state is represented, relative to the energy eigenbasis, by a matrix with elements $\rho_{jk}(t)  \coloneqq  \bra{j}  \rho_\Sys(t)  \ket{k}$,
wherein $j, k \in \{0, 1\}$.
The matrix elements evolve under 
a complex \emph{decoherence function} $G(t)$:
\begin{align}
   \label{eq_Matrix_coheres}
   \rho_\Sys(t) = \begin{bmatrix}
   1 - | G(t) |^2  \,  \rho_{11}(0)  &  G(t) \, \rho_{01}(0)  \\
   G(t)  \,  \rho_{10}(0)  &  | G(t) |^2  \,  \rho_{11}(0)
   \end{bmatrix} .
\end{align}
To specify $G(t)$, we assume that $\Env$ has a Lorentzian spectral density function (SDF) of width $\lambda$. Define the function
$\lambda'  \coloneqq  \sqrt{ \lambda^2  -  2 \gamma_0 \lambda } \, ,$
whose $\gamma_0$ depends on the couplings $g_\ell$.
Suppose that the mode is on resonance with the qubit: The SDF is centered at $\Delta$. The decoherence function becomes
\begin{align}
   G(t)
   =  e^{- \lambda t / 2}  
   \left[  \cosh  \left( \frac{\lambda' t}{2}  \right)
   +  \frac{\lambda}{\lambda'}  \,  \sinh  \left(  \frac{\lambda' t}{2}  \right)  \right] .
\end{align}
If $\gamma_0  <  \lambda / 2$, the coupling is weak, and the bath is Markovian.
$G(t)$ is real and decreases monotonically with $t$.
If $\gamma_0  >  \lambda / 2$, the coupling is strong, and the bath is non-Markovian.

\begin{figure}[h]
\centering
\begin{subfigure}{0.4\textwidth}
\centering
\includegraphics[width=1\textwidth]{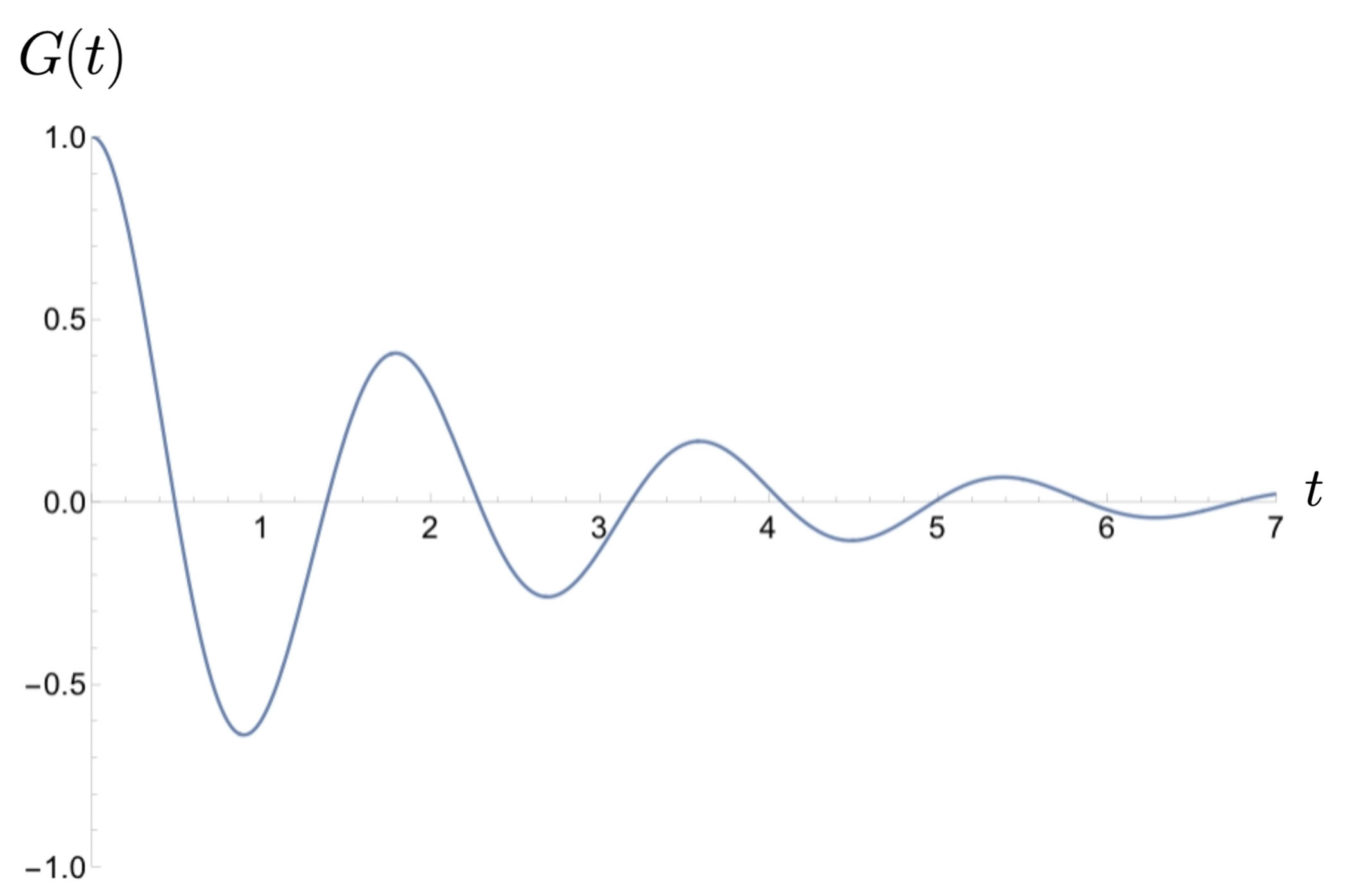}
\caption{\caphead{Decoherence function vs. time.}}
\label{fig_NonMarkov_Coher}
\end{subfigure}
\begin{subfigure}{0.4\textwidth}
\centering
\includegraphics[width=1\textwidth]{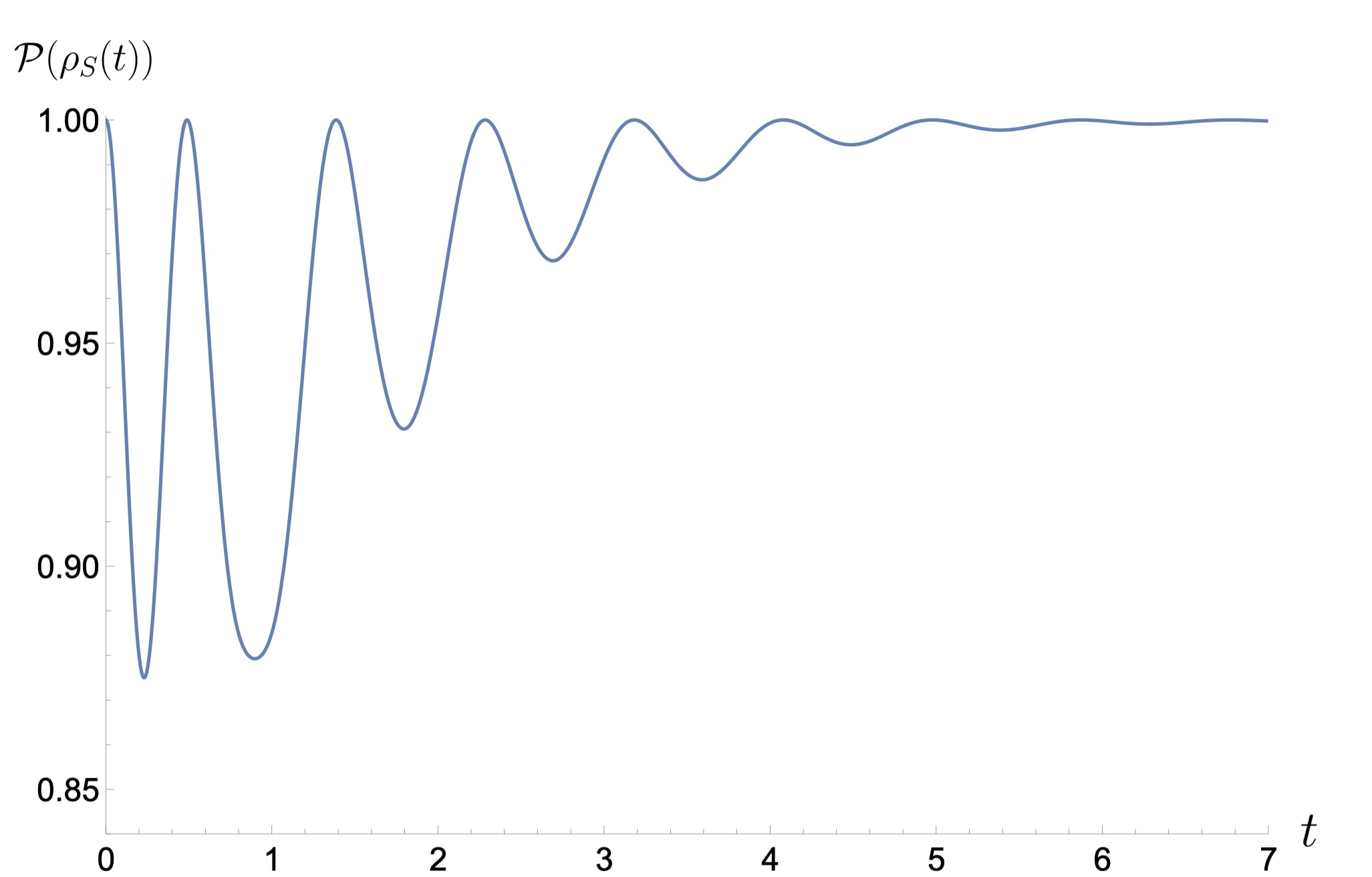}
\caption{\caphead{Purity vs. time.}}
\label{fig_NonMarkov_Purity}
\end{subfigure}
\caption{\caphead{Coherence and purity at strong coupling:} 
A bosonic mode decoheres a qubit initialized in $(\ket{0} + \ket{1}) / \sqrt{2}$.
The Lorentzian width $\lambda = 1$. The coupling $\gamma_0 = 50 \lambda / 2$ is strong, so the bath is non-Markovian. 
For all $t$, the decoherence function $G(t) \in \mathbb{R}$.  $G(t)$ and the purity, $\pure \LParen  \rho_\Sys(t)  \RParen$, oscillate. Hence non-Markovianity can revive the purity. ($\pure$ eventually reaches 1 because the evolution maps all states to $\ket{0}$.)
}
\label{fig_NonMarkov}
\end{figure}

\begin{figure}[h]
\centering
\begin{subfigure}{0.4\textwidth}
\centering
\includegraphics[width=1\textwidth]{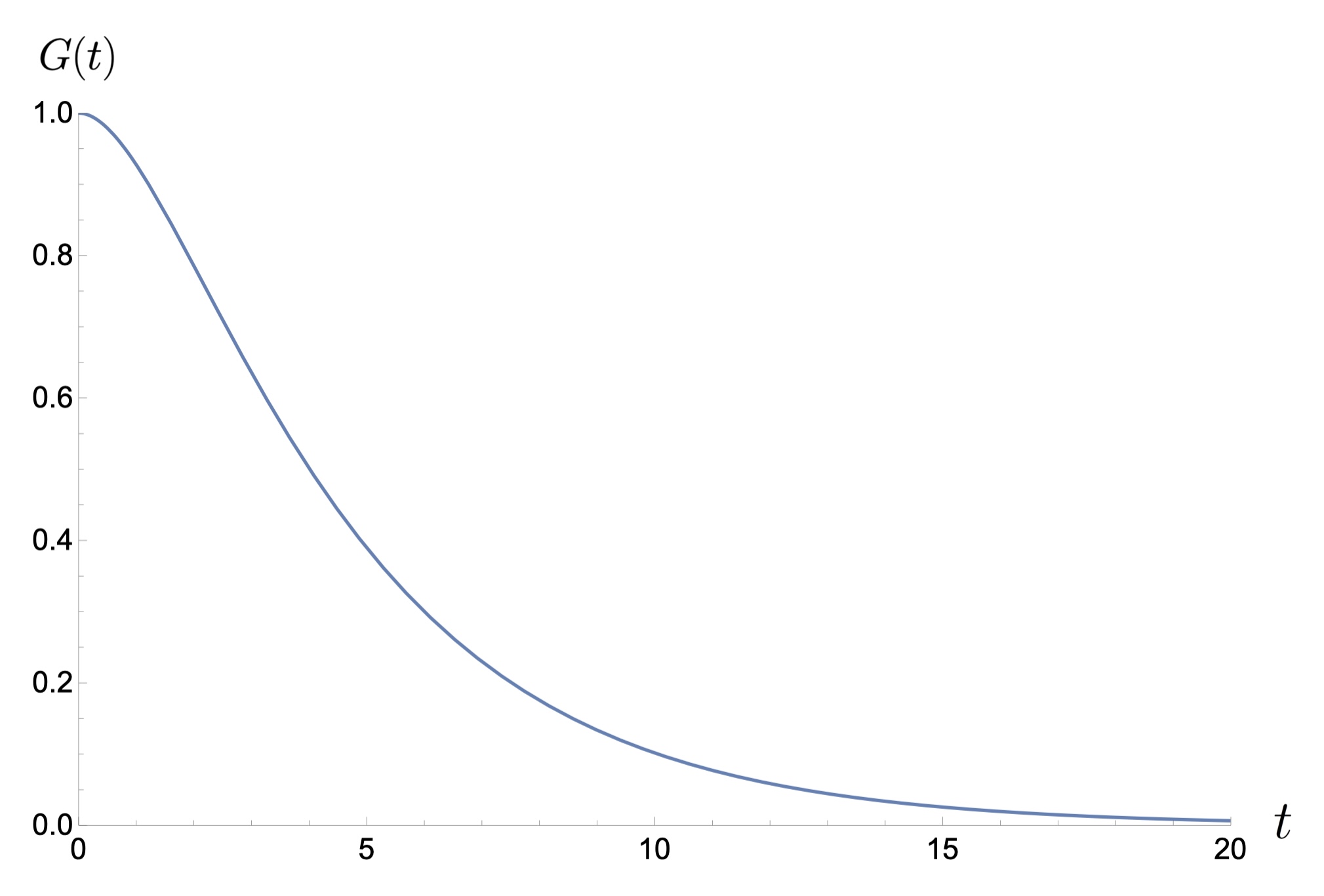}
\caption{\caphead{Decoherence function vs. time.}}
\label{fig_Markov_Coher}
\end{subfigure}
\begin{subfigure}{0.4\textwidth}
\centering
\includegraphics[width=1\textwidth]{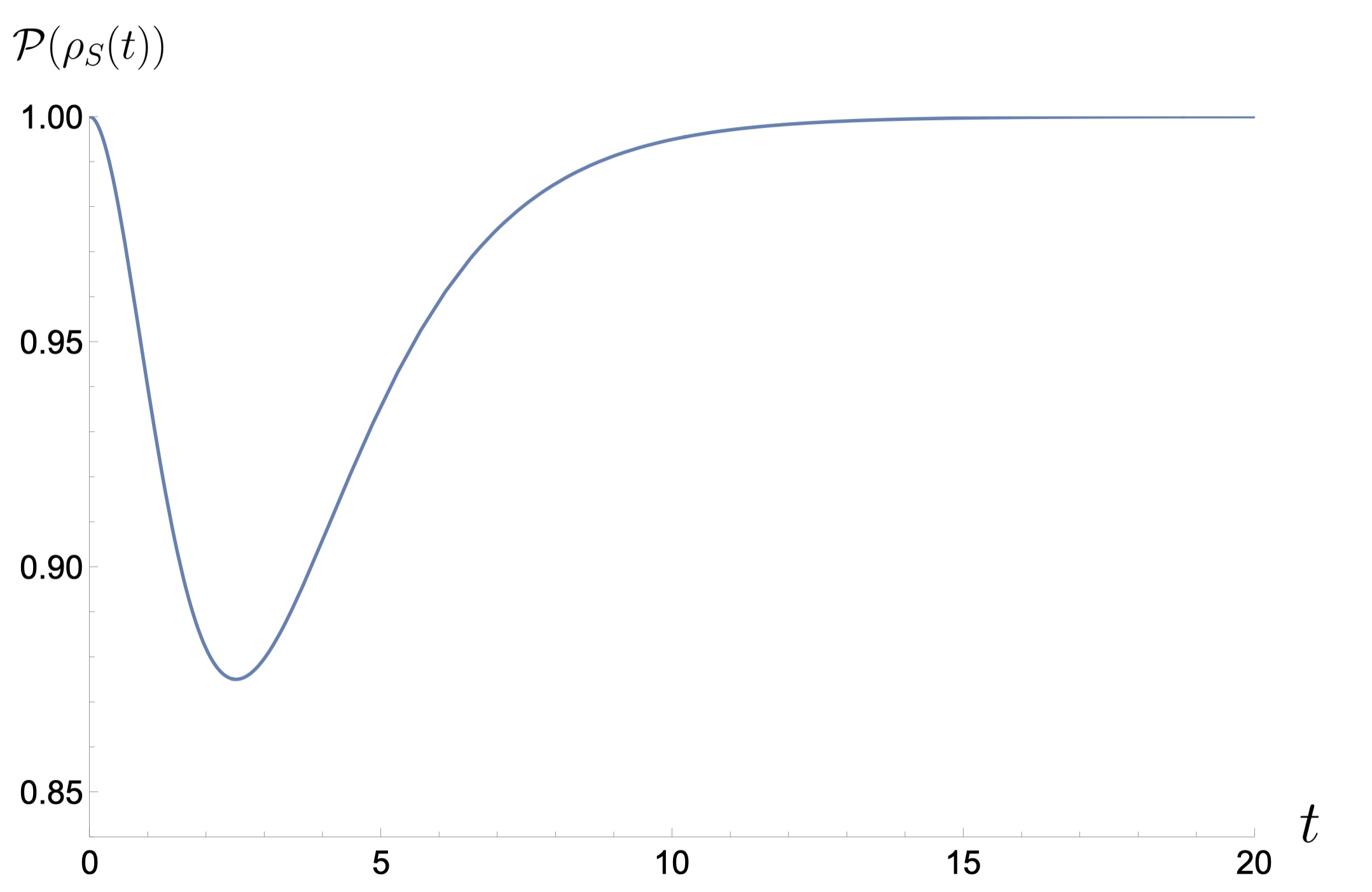}
\caption{\caphead{Purity vs. time.}}
\label{fig_Markov_Purity}
\end{subfigure}
\caption{\caphead{Coherence and purity at weak coupling:} 
The setup is mostly as for Fig.~\ref{fig_NonMarkov_Purity}. However, the coupling $\gamma_0 = 0.4 \lambda$ is weak. Unlike in Fig.~\ref{fig_NonMarkov_Purity}, $G(t)$ and $\pure$ do not oscillate.
}
\label{fig_Markov}
\end{figure}

Figure~\ref{fig_NonMarkov} illustrates effects of a non-Markovian bath, whose coupling is strong: $\gamma_0 = 50 \lambda / 2$.
Figure~\ref{fig_NonMarkov_Coher} shows the decoherence function vs. time. $G(t)$ oscillates, enabling the qubit's energy coherences to revive repeatedly. The oscillations' envelope decays eventually, however, demonstrating a limitation of non-Markovianity as a tool. Figure~\ref{fig_NonMarkov_Purity} shows the purity $\pure \LParen \rho_{\Sys} (t) \RParen = \Tr \LParen \rho_\Sys(t)^2 \RParen$ of a qubit state initialized to $(\ket{0} + \ket{1}) / \sqrt{2}$. Calculated from Eqs.~\eqref{eq_Matrix_coheres}, the purity oscillates until asymptoting. $\pure$ asymptotes because $\Sys$ ends up in $\ket{0}$, as the environment began in its vacuum state and so sucks the energy out of $\Sys$. However, the oscillations demonstrate that the environment's non-Markovianity partially returns $\Sys$ to its initial state repeatedly.
Figure~\ref{fig_Markov} depicts the same quantities---$G(t)$ and the purity---in the absence of non-Markovianity: The coupling $\gamma_0 = 0.4 \lambda$ is weak. Neither function oscillates.
Hence non-Markovianity can help baths achieve the energy criterion~\ref{crit_Energy} without threatening the purity criterion~\ref{crit_Purity} as much as thermal baths do, for a time.

%

\section{Two topics adjacent to the criteria}
\label{sec_Exclusions}

This section describes the relationship between our criteria and each of two adjacent topics: instructions and measurements. Multiple of our criteria, together, imply that useful autonomous quantum machines obey instructions (Sec.~\ref{sec_Instructions}). Furthermore, useful autonomous quantum machines do not generally require measurements (Sec.~\ref{sec_Measurement}).

\subsection{Instructions}
\label{sec_Instructions}

Certain autonomous quantum machines appear to call for a ninth criterion: instructions. Instructions could guide an autonomous quantum drone to walk some distance forward, turn leftward, and then turn off. In another example, an autonomous quantum computation should implement one algorithm, rather than another~\cite{Meier_24_Autonomous}. Upon satisfying our criteria, however, one can effect instructions, which form a kind of emergent criterion. In the first example, one could guide the drone with a track covered by the mobility criterion (\ref{crit_Mobility}). The switching-off criterion (\ref{crit_Off}) would enforce the final instruction. In the second example, one can feed an autonomous quantum computation instructions by leveraging a clock (criterion~\ref{crit_Time}), interactions (criterion~\ref{crit_Interact}), interoperability (criterion~\ref{crit_Interop}), etc. Single-purpose machines, such as autonomous quantum engines, do not require such instructions.

\subsection{Measurements}
\label{sec_Measurement}

One might expect our criteria to include measurement, for two reasons. First, an agent needs confidence in an autonomous quantum machine's ability to accomplish its mission. One might gain such confidence by measuring the machine's output. The need for confidence, however, implies only the need to test the machine before deployment. The testing need not be autonomous; only the machine's operation need be autonomous. Once the agent attains confidence in the machine's operation, the testing---the nonautonomous activity---can cease.

Second, typical quantum experiments end with measurements~\cite{NielsenC10}. An autonomous quantum machine's operation may appear to constitute a quantum experiment. Hence one may expect to end such a machine's operation with a measurement. However, a typical quantum experiment is intended to furnish information about the natural world or an artificial system---to answer questions such as \emph{does a particular system exhibit a phase transition? How quickly does a particular qubit decohere? Which path does a particle traverse?} Autonomous quantum machines are not generally intended to answer such questions; they serve purposes different from typical quantum experiments. Quantum refrigerators are intended to cool, quantum engines are intended to provide work, molecular switches can be intended to store energy~\cite{Kucharski_14_Templated}, etc. Hence autonomous quantum machines generally need not follow all the requirements, including measurements, of typical quantum experiments.

One might object that three autonomous quantum machines provide information about natural and artificial systems, like typical quantum experiments: autonomous quantum clocks, sensors, and computers. After all, the clocks described in Sec.~\ref{crit_Time} answer the question \emph{at which point has a predetermined time interval passed since the last tick?} Sensors report overtly about their environments. Quantum computers help answer questions such as \emph{what are the prime factors of a particular number?} Still, these three examples do not imply that autonomous quantum machines generally require measurements. The clocks of Sec.~\ref{crit_Time} do not keep time not for macroscopic agents who obtain information about quantum systems only through measurements. Rather, the clocks keep time for other quantum systems, by emitting excitations. An open-quantum-systems framework describes such interactions effectively~\cite{Erker_17_Autonomous}. Therefore, positive-operator-valued measures (POVMs)---mathematical representations of quantum measurements~\cite{NielsenC10}---are unnecessary and arguably inefficient as descriptions. Autonomous quantum sensors may emit excitations similarly to announce their findings. Hence autonomous quantum clocks and sensors do not necessarily require conventional measurements. Furthermore, criteria~\ref{crit_Time} and~\ref{crit_Output} (timekeeping mechanisms and output worth the input) cover the demands on these devices' outputs; an extra measurement criterion would be redundant. Finally, even if clocks and sensors required conventional measurements, they would not justify a measurement criterion, which would be irrelevant to many other autonomous quantum machines.

Quantum computation does require conventional measurements, according to DiVincenzo's criteria~\cite{DiVincenzo_00_Physical}. Let us assume, for the sake of argument, that conventional measurements require classical intervention, or are not autonomous. Fully autonomous quantum computers appear impossible---not due to impracticality, but because they are oxymorons. Still, partially autonomous quantum machines can be useful, as detailed in item~\ref{item_Motivate_auto} of Sec.~\ref{sec_Outlook}. Hence a quantum computer that undergoes an autonomous state preparation or an autonomous circuit, followed by a nonautonomous measurement, can be useful. The preparation's or circuit's output (criteria~\ref{crit_CPU} and~\ref{crit_Output}) would be a quantum state. Therefore, no measurement would play any role in the autonomous quantum machine's operation; a measurement would only follow that operation. For consistency with this point, we often write \emph{autonomous quantum circuit,} rather than \emph{autonomous quantum computer,} in the main text.

In summary, useful autonomous quantum machines do not require measurements generally. For example, once a quantum refrigerator has cooled an ancilla qubit to be used in a quantum computation, the ancilla should simply be used. It need not be measured, if the quantum refrigerator has withstood sufficient testing. Autonomous quantum clocks and sensors do need to communicate with the classical world. Still, POVMs do not always represent this need best, and other criteria cover the need. Quantum computers, requiring measurements, cannot be fully autonomous. Therefore, research should focus on partially autonomous quantum computation---autonomous state preparation and autonomous quantum circuits.

\end{appendices}

%
%
\bibliography{AQM_refs}


\end{document}